# A Biomechanical Study on the Use of Curved Drilling Technique for Treatment of Osteonecrosis of Femoral Head

Mahsan Bakhtiarinejad[1], Farshid Alambeigi[1], Alireza Chamani[1], Mathias Unberath[2], Harpal Khanuja[3], Mehran Armand[1,3,4]

[1]Laboratory for Computational Sensing and Robotics, Department of Mechanical Engineering, Johns Hopkins University, Baltimore, MD, USA, 21218.

[2]Laboratory for Computational Sensing and Robotics, Department of Computer Science, Johns Hopkins University, Baltimore, MD, USA, 21218.

[3]Department of Orthopedic Surgery, Johns Hopkins Medical School, Baltimore, MD, USA, 21205.

[4]Johns Hopkins University, Applied Physics Laboratory, Laurel, MD, USA, 20723.

**Abstract**

Osteonecrosis occurs due to the loss of blood supply to the bone, leading to spontaneous death of the trabecular bone. Delayed treatment of the involved patients results in collapse of the femoral head, which leads to a need for total hip arthroplasty surgery. Core decompression, as the most popular technique for treatment of the osteonecrosis, includes removal of the lesion area by drilling a straight tunnel to the lesion, debriding the dead bone and replacing it with bone substitutes. However, there are two drawbacks for this treatment method. First, due to the rigidity of the instruments currently used during core decompression, lesions cannot be completely removed and/or excessive healthy bone may also be removed with the lesion. Second, the use of bone substitutes, despite its biocompatibility and osteoconductivity, may not provide sufficient mechanical strength and support for the bone. To address these shortcomings, a novel robot-assisted curved core decompression (CCD) technique is introduced to provide surgeons with direct access to the lesions causing minimal damage to the healthy bone. In this study, with the aid of finite element (FE) simulations, we investigate biomechanical performance of core decompression using the curved drilling technique in the presence of normal gait loading. In this regard, we compare the result of the CCD using bone substitutes and flexible implants with other conventional core decompression techniques. The study finding shows that the maximum principal stress occurring at the superior domain of the neck is smaller in the CCD techniques (i.e. 52.847 MPa) compared to the other core decompression methods; furthermore, the peak value of normal stress at the interface for the CCD model is substantially smaller than traditional and advanced core decompression techniques (89% and 76%, respectively). FE results demonstrate the superior performance of CCD compared to the other approaches without any compromise to patient's safety and markedly reducing the risk of femoral fracture in the postoperative phase for normal gait loading.

## 1 Introduction

Avascular necrosis (AVN) of the femoral head; also known as osteonecrosis, is a degenerative joint disease, which happens due to the interruption of blood supply to the bone leading to the spontaneous death of the trabecular bone (Fig. 1a and b) [1, 2]. The rate of occurrence of this disease in the United states is between 10,000-30,000 per year, which typically happens in young adults with around 35-50 years of age. Studies show that due to late diagnosis and treatment, more than 85% of the early stage AVN leads to collapse of the femur's subchondral and finally total hip arthroplasty [2]. The key parameter in successful treatment of AVN is complete removal of the necrotic region with individual variation in shapes and locations, which is typically not easily accessible due to the complexity of the femur anatomy and utilizing rigid instruments [3].

Literature documents several early-stage (i.e. before femoral head collapse) AVN treatments including osteotomy [4] and various types of core decompressions [5]. Among these approaches, core decompression (CD) is a well-established procedure performed with the goal of reducing the femoral head pressure and revitalizing the bone by restoring blood flow to the lesion area. In the simple CD procedure, surgeon drills either one straight tunnel of 8-10 mm diameter or multiple tunnels of 3.2 mm diameter with a single-entry point in the lateral subtrochanteric region [2].
This approach is plagued with the following challenges:
(i) The rigidity of the tools currently utilized do not provide the surgeon with sufficient access for complete removal of necrotic lesions; each of which tend to be of a unique and amorphous shape.
(ii) Consequential weakening of the healthy bone puts the patient at risk of subchondral bone collapse either during or after surgery, due to the patient's weight. As a remedy to the second issue, some surgeons use structural bone grafts [6] or bone grafts with porous tantalum implant [7] to provide additional structural support for the subchondral bone and stimulate bone formation with blood and bone cells.

To address the accessibility and the structural strength issues, Landgraeber et al. [8] developed the advanced core decompression (ACD) technique, that uses a percutaneous expandable reamer together with an artificial bone graft (i.e. PRO-DENSE ®, Wright Medical TM, Arlington, TN, USA). Consequently, they could show a better accessibility to the necrotic lesion area and achieve a relatively good compressive strength for the first 13 weeks after surgery. However, this approach has two main shortcomings. First, in order to reach the necrotic region, the expandable reamer generates a relatively large conical hole, which unnecessarily removes healthy tissues along with the necrotic lesions, thereby potentially further weakening the femoral head. Also, the incorporated bone substitute becomes weaker after 26 weeks due to triphasic resorption [9].

To address the aforementioned challenges in core decompression, curved drilling technique has been recently proposed [10, 11]. In this approach, authors used a continuum dexterous manipulator [12,13] and developed appropriate flexible drilling tools [14,15] to increase the reachability into necrotic regions while minimizing the damage to safe tissues as opposed to the mentioned techniques utilizing rigid instruments. In this study, we investigate the effect of curved core decompression (CCD) on the structural stability of the femoral head after AVN treatment using biomechanical analysis.

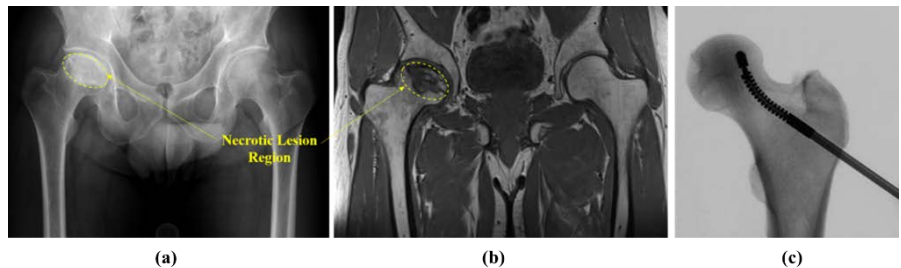

**Fig. 1:** (a) Fluroscopic and (b) MRI image of a patient suffering from osteonecrosis of right femoral head (Case courtesy of Dr Henry Knipe, Radiopaedia.org, rID: 44260); (c) Fluroscopic image of the core decompression procedure using the curved drilling technique.

We perform Finite Element (FE) analysis to evaluate the biomechanical performance of the CCD (Fig. 1c) against (i) traditional core decompression and (ii) ACD technique reinforced using PRO-DENSE bone substitute. In order to evaluate neck stability and fracture risk after core decompression, we also investigate traditional CD and CCD techniques using flexible Nitinol implant as an alternative bone substitute.

## 2 Materials and Methods

### 2.1 Geometrical Modeling

We first generate a three-dimensional (3D) model of the femur from segmented Computed Tomography (CT) scan of the specimen. Through a threshold-based segmentation of the bony anatomy, the domains of the femur are identified using image processing software; Medical Imaging Interaction Toolkit (MITK) [16]. Manual refinement improves the segmentation and smoothens the three-dimensional femur model [17]. Using SolidWorks (Dassault Systèmes SolidWorks Corporation, MA, USA) software, we first consider an arbitrary necrotic region area, which is not completely accessible with the simple CD approach and then, modify the 3D models based on the ACD and CCD techniques. Fig. 2 demonstrates aforementioned 3D models together with the considered drilled tunnels and an arbitrary lesion area in the femoral head. For all models, we consider a similar single-entry point in the lateral subtrochanteric region and lesion area to study the biomechanical performance of these approaches in the presence of identical normal walking loads. For the case of ACD, the size of the conical head is designed such that the tunnel can reach the lesion region.

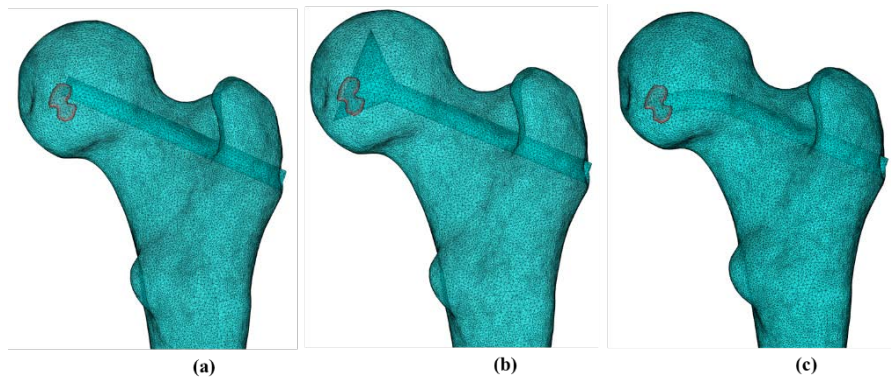

**Fig. 2:** Geometrical and finite element models of (a) CD, (b) ACD and (c) CCD configurations with the arbitrary lesion area

### 2.2 Finite Element Simulation

The finite element model of the femur is generated using linear 10-node tetrahedral elements (Fig. 2) in COMSOL Multiphysics® (COMSOL AB,

Stockholm, Sweden). An automated mesh convergence is performed during the simulation in COMSOL and the largest mesh size that would produce the converging results is chosen. The element numbers for final meshes are displayed in Table 1.

**Table 1:** Number of the elements in FE model meshes

| Model | Element |
|---|---|
| CD | 524940 |
| ACD | 532832 |
| CCD | 524894 |
| Untreated | 495062 |

For material properties of the femur, a linear, elastic and isotropic bone with 3 mm uniform thickness of the cortical bone layer [18] and internal cancellous bone is considered. For this purpose, the distance between triangular surfaces of the femur and centroids of the bone elements are found using closest neighbor and bounding spheres approach of the ICP algorithm [19]. The elements within 3 mm of the surface bone are assigned as cortical and the remaining as cancellous. The average Young's modulus and Poisson's ratio of the cortical and cancellous regions of the bone are considered as 12 GPa and 600 MPa and 0.40 and 0.29, respectively. PRO-DENSE and Nitinol elements are also assumed linear, elastic and isotropic with average Young's modulus of 221.6 MPa and 83 GPa and Poisson's ratio of 0.3 and 0.33, respectively [2,18, 20, 21].

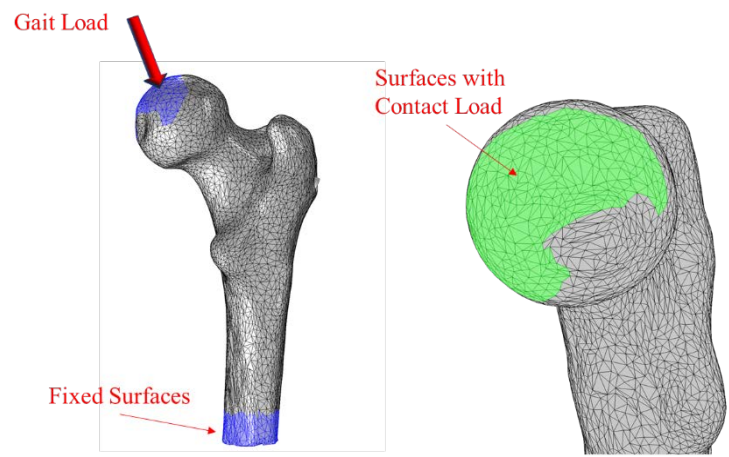

**Fig. 3:** Fixed boundary and loading condition

We investigate the fracture risk for the peak hip joint loads during normal walking. The applied load of 2132.8 (-411.24, -247.1, 2078.14) N is extracted from the patient data set "HSR" measured during normal walking [18, 22]. The load is distributed over the cartilaginous surface of the femur that is in contact within the hip joint. A fixed boundary condition is applied at the distal end of the femur (Fig. 3).

## 3 Computational Results

In this study, we define the bone failure criteria based on the maximum principal stress and maximum principal strain. During normal walking, tensile stress emerges in the subtrochanteric region; therefore, maximum first principal stress of the femur

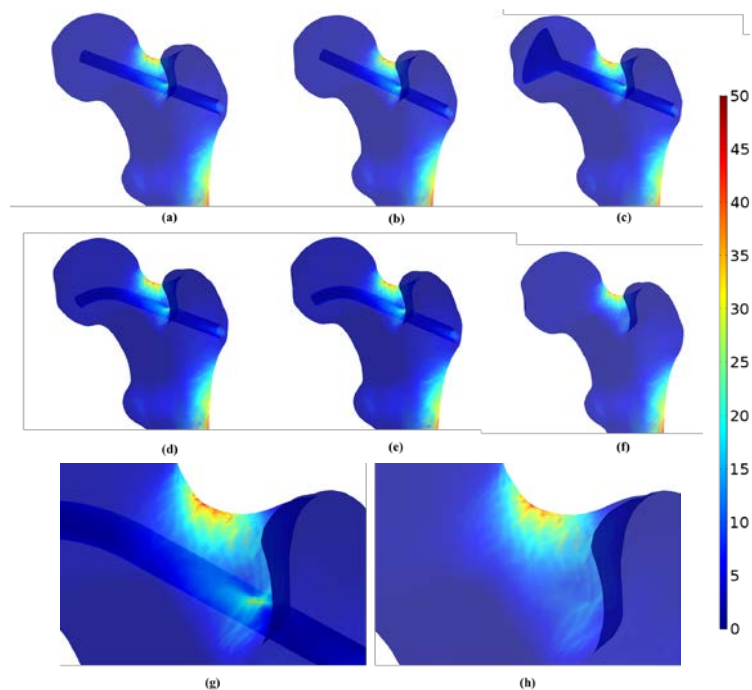

**Fig. 4:** First Principal stress distribution (MPa) in femoral bone using (a) CD with PRO-DENSE, (b) CD with Nitinol (c) ACD with PRO-DENSE, (d) CCD with PRO-DENSE, (e) CCD with Nitinol and (f) Untreated model. (g) Enlarged view of CCD with PRO-DENSE (h) Enlarged view of Untreated model

is compared with $\sigma_{ult}^t$=110 MPa which is the tensile strength of cortical bone reported in the literature [2,18]. The maximum first principal stress given in percent of the tensile strength are displayed in Table 2 for all models.

**Table 2:** Maximum first principal stress of the femur in percent of the tensile strength $\sigma_{ult}^t$ subject to normal walking loading

| Model | Maximum First Principal stress (MPa) | Percent of $\sigma_{ult}^t$ |
|---|---|---|
| CD with PRO-DENSE | 53.678 | 49 |
| CD with Nitinol | 53.676 | 49 |
| ACD with PRO-DENSE | 55.667 | 51 |
| CCD with PRO-DENSE | 52.856 | 48 |
| CCD with Nitinol | 52.847 | 48 |
| Untreated | 51.872 | 47 |

All the treated configurations and untreated femur exhibit typical tensile stress distribution subject to gait loading as shown in Fig. 4. The maximum principle stress in all models occurs at the superior domain of the neck and is smaller than tensile strength $\sigma_{ult}^t$. Furthermore, high stresses emerge in distal regions of the subtrochanter as well as in the small regions around the drilling areas within the bone (Fig. 4g and h).

In the treated models, the maximum tensile stress is 48-51% of the appropriate tensile strength which is slightly larger than that of the untreated model (47%). Therefore; the fracture risk is estimated quite low for all the models subject to normal walking loading. The peak normal and shear stresses at the interface for all treated models are also computed and shown in Table 3. The peak stress values of the CD and CCD configurations decrease substantially using a bone substitute with lower modulus of elasticity; consequently, using stiffer filling material; Nitinol, increasingly raises the peak of the normal and shear stresses compared to the PRO-DENSE.

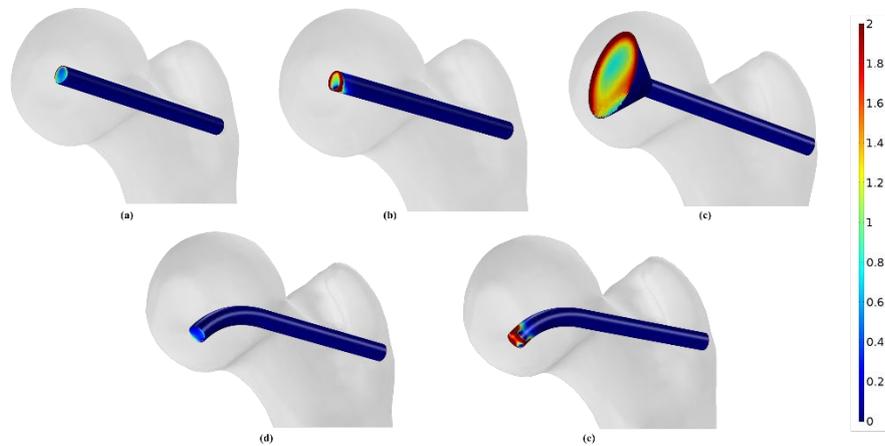

**Fig. 5:** Normal stress distribution (MPa) at the interface using (a) CD with PRO-DENSE, (b) CD with Nitinol (c) ACD with PRO-DENSE, (d) CCD with PRO-DENSE and (e) CCD with Nitinol

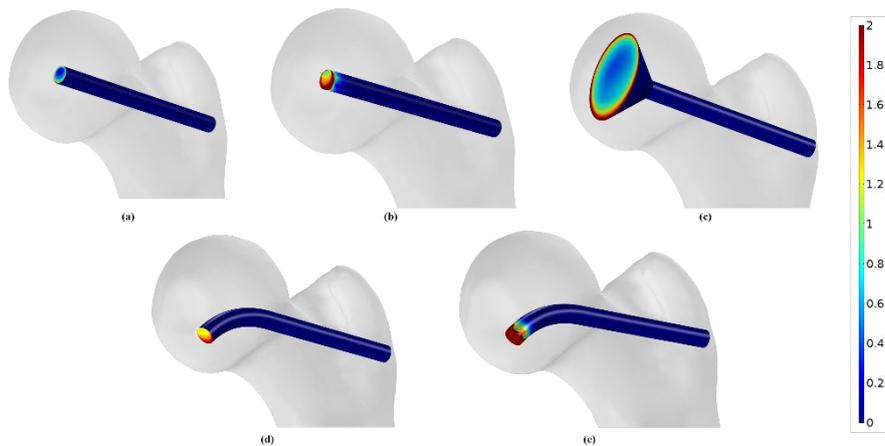

**Fig. 6:** Shear stress distribution (MPa) at the interface using (a) CD with PRO-DENSE, (b) CD with Nitinol (c) ACD with PRO-DENSE, (d) CCD with PRO-DENSE and (e) CCD with Nitinol

Similar distribution of normal and shear stresses is observed at the interface of the treated models (Fig. 5 and 6) with the peak values appearing at the tip of the interface located in the femoral head. The peak value of the normal stress for the CCD technique using PRO-DENSE is 1.96 MPa which is substantially lower than that of the CD and ACD (4.549 and 8.058 MPa). In addition, the shear stress of CCD using PRO-DENSE has a peak value of 4.39 MPa which is lower than that of the ACD (5.941 MPa); however, it is higher than peak shear stress for the CD procedure (2.61 MPa). The distribution of normal and shear stress of the ACD configuration using PRO-DENSE demonstrates that due to larger surfaces of the tip of the tunnel compared to the CCD procedure, surfaces of high stress at the femoral head are more spread.

**Table 3:** Peak normal stress and shear stress at the interface

| Model | Peak Normal Stress (MPa) | Peak Shear Stress (MPa) |
|---|---|---|
| CD with PRO-DENSE | 4.549 | 2.606 |
| CD with Nitinol | 17.368 | 14.356 |
| ACD with PRO-DENSE | 8.058 | 5.941 |
| CCD with PRO-DENSE | 1.958 | 4.388 |
| CCD with Nitinol | 45.428 | 40.141 |

Based on the principal maximum strain criterion, the maximum ($\varepsilon_{max}$) and minimum ($\varepsilon_{min}$) principal strain values computed at the centroid of each bone element, are used to determine the volume of the failed elements where the bone tissue reaches the yield strain. For this purpose, greater value of $|\varepsilon_{max}|$ and $|\varepsilon_{min}|$ is compared with the appropriate compressive and tensile yield strains which is $\varepsilon_{yC}$ =0.0104 for compression and $\varepsilon_{yT}$ =0.0073 for tension [23,24]. Consequently, the volume of the failed elements is computed by adding up the volume of the elements that have a strain higher than its yield value (Table 4). It is demonstrated that all treated and untreated models have similar percentage of the volume failure, that is 0.04% of the total bone volume.

**Table 4:** Volume of the failed elements using principal maximum strain criterion

| Model | Volume of Failed Elements (mm$^3$) | Femur Volume (mm$^3$) |
|---|---|---|
| CD with PRO-DENSE | 56 | 141483 |
| CD with Nitinol | 56 | 141483 |
| ACD with PRO-DENSE | 55 | 138687 |
| CCD with PRO-DENSE | 57 | 141416 |
| CCD with Nitinol | 57 | 141416 |
| Untreated | 55 | 143419 |

## 4 Discussion and Conclusion

The ultimate tensile stress of the cortical bone reported in the literature varies from 78 MPa to 151 MPa [18, 25-27]. In this study, a limit tensile strength of 110 MPa is chosen as a reference [2, 9, 18]. According to Table 2, the maximum principal stress in untreated and treated models is on average about 52 MPa, which is smaller than both the ultimate tensile strength of the chosen cortical bone (i.e. 110 MPa) and the reported maximum principal stress (i.e. 75.46 MPa) [28]. It is demonstrated that the maximum principal stress in the treated models using curved drilling technique utilizing PRO-DENSE (i.e. ~52.847 MPa) is smaller than the other techniques (i.e. ACD and CD). Moreover, the peak value of normal stress at the interface for the CCD model using PRO-DENSE decreases substantially when compared to the CD and ACD (57% and 76 %, respectively); in addition, the peak value of shear stress for the CCD with PRO-DENSE is 26% lower than ACD showing the advantage of CCD compared to the other approaches.

Simulation results; therefore, confirm that the CCD approach like the other approaches does not increase the risk of femoral fracture. Tren et. al [9] also achieved similar results for their ACD technique. However, ACD unnecessarily removes healthy tissues and results in larger values of maximum principal stress at the superior domain of the femoral neck and peak values of normal and shear stresses at the tip of the interface compared to CCD (Tables 2 and 3).

The simulation of CCD technique with PRO-DENSE generates the smallest peak value of normal stress; whereas, CCD with Nitinol results in the largest peak stress. This result is in accordance with the reported results in [7] and [9] representing that the use of a stiff implant increases the maximum stress values whilst incorporating a lower stiffness of the bone substitutes (e.g. PRO-DENSE) may have better outcomes. On the other hand, as mentioned in [9], incorporating bone substitute with low stiffness such as PRO-DENSE may increase the risk of femoral fracture due to the triphasic resorption after 26 weeks [9]. Considering this limitation and the acceptable maximum principal stress of the bone using CCD with Nitinol implant, the use of this implant can be justified as compared to the PRO-DENSE. A potential path forward that may constitute a healthy compromise between the two aforementioned approaches can be the use of a calcium phosphate bone cement reinforced with a bendable implant. This approach will increase bone strength while minimizing femoral damage risk during the first weeks of the post-operative phase.

As shown in Table 3, 4 and Fig. 5 and 6, the maximum principal, normal and shear stresses of ACD undergo larger values compared to the CCD method. ACD, therefore, may compromise safety by removing the healthy bone tissue and increasing the risk of collapse of the articular surface [29, 30]. It also may cause femoral head depression during and after the operation which happens due to larger surfaces of high stress at the tip of the interface located at the femoral head.

This preliminary study is performed on a simulated lesion on a femur model. Future studies must consider the effect of each treatment technique in the presence of a variety of lesion geometries. Also, in the future, the FE model of the CCD technique using PRO-DENSE in combination with the Nitinol implant need to be investigated and further enhanced to include more geometric details including a more accurate interface between the bone and the implant. Optimizing the bending curvature when approaching the lesion location may also enhance the effectiveness of the CCD approach. Eventually, the FE model analysis can become a part of the AVN evaluation and prognosis procedures, providing patient-specific results.

**Acknowledgments**

Research reported in this publication was supported by the National Institute of Arthritis and Musculoskeletal and Skin Diseases of the National Institutes of Health under Award Number R01EB016703 and R01EB023939. The content is solely the responsibility of the authors and does not necessarily represent the official views of the National Institutes of Health.